\newcommand{\HII}{\mbox{H\hspace{0.2em}{\scriptsize II}}}

\newcommand{\CII}{\mbox{[C\hspace{0.2em}{\scriptsize II}]}}
\newcommand{\al}{\mbox{$^{26}$\hspace{-0.2em}Al}}
\newcommand{\Msol}{M_{\sun}}

\newcommand{\gray}{\mbox{$\gamma$-ray}}

\newcommand{\um}{\mbox{$\mu$m}}

\def\phibar{\bar{\varphi}}

\def\MeV{\mbox{Me\hspace{-0.1em}V}}

\def\deg{\hbox{$^\circ$}}
\def\sun{\hbox{$\odot$}}

\def\apj{ApJ}	
	
\def\apjs{ApJS}

\def\aap{A\&A}	
		
\def\aaps{A\&AS}

\def\mnras{MNRAS}

\def   \ni {\noindent}

\def   \ssk {\vskip  5truept}

\def   \bsk {\vskip 15truept}

\def   \newpage {\vfill\eject}
\def   \newline {\hfil\break}

\documentstyle[epsfig]{article}
\begin{document}
\hsize 5truein
\vsize 8truein
\font\abstract=cmr8
\font\keywords=cmr8
\font\caption=cmr8
\font\references=cmr8
\font\text=cmr10
\font\affiliation=cmssi10
\font\author=cmss10
\font\mc=cmss8
\font\title=cmssbx10 scaled\magstep2
\font\alcit=cmti7 scaled\magstephalf
\font\alcin=cmr6
\font\ita=cmti8
\font\mma=cmr8
\def\ref{\par\noindent\hangindent 15pt}
\null

\title{\ni Assessment of Tracers of 1.8 \MeV\ Emission}

\bsk \bsk
\author{\ni J.~Kn\"odlseder$^1$,
            R.~Diehl$^2$,
            U.~Oberlack$^5$,
	        P.~von Ballmoos$^1$,
            H.~Bloemen$^3$,
            W.~Hermsen$^3$, \\
            A.~Iyudin$^2$,
	        J.~Ryan$^4$,
	        and
	        V.~Sch\"onfelder$^2$}
\bsk
\affiliation{$^1$Centre d'Etude Spatiale des Rayonnements, CNRS/UPS, B.P.~4346,
	         31028 Toulouse Cedex 4, France}
\affiliation{$^2$Max-Planck-Institut f\"ur extraterrestrische Physik,
	         Postfach 1603, 85740 Garching, Germany}
\affiliation{$^3$SRON-Utrecht, Sorbonnelaan 2, 3584 CA Utrecht, The
	         Netherlands}
\affiliation{$^4$Space Science Center, University of New Hampshire, Durham
	         NH 03824, U.S.A.}
\affiliation{$^5$Astrophysics Laboratory, Columbia University, New York, 
	         NY 10027, U.S.A.}
\bsk
\baselineskip = 12pt

\abstract{ABSTRACT \ni 
We examined the question of a possible correlation between 1.8 \MeV\ emission
and intensity distributions observed at other wavelengths by comparing the 
1.8 \MeV\ data to an extended set of all-sky maps, covering the entire explored
wavelength range from the radio band up to high-energy gamma-rays. 
This analysis revealed that tracers of the old stellar population or the local
interstellar medium provide only a poor description of the data. 
Tracers of the young stellar population considerably improve the fit. 
Residuals are minimized for the free-free emission map obtained by the DMR 
at microwave wavelengths and the 158 $\mu$m fine-structure line map of 
C$^+$ observed by FIRAS.
Within the statistics of the present data, both maps provide an 
overall satisfactory fit.
We therefore conclude that adequate tracers of \al\ sources have now 
been identified.
Implications for \al\ source types are discussed
(see also Kn\"odlseder, these proceedings).
}
\bsk
\baselineskip = 12pt
\keywords{\ni KEYWORDS: gamma-ray lines; 
                        nucleosynthesis;
                        massive stars;
                        multi-wavelength analysis;
}

\bsk
\baselineskip = 12pt

\text{
\ni 1. INTRODUCTION
\ssk
\ni
\label{sec:intro}

The imaging telescope COMPTEL aboard the {\em Compton Gamma-Ray 
Observatory} ({\em CGRO}) provided the first all-sky map of the sky 
in the light of the 1.809 \MeV\ \gray\ line, attributed to the 
radioactive decay of \al\ (Oberlack et al.~1996; Kn\"odlseder et 
al.~these proceedings).
This map shows an asymmetric emission profile along the galactic 
plane with peculiar features towards the direction of Cygnus, Vela, 
Carina and near the anticenter.
It has been shown that this type of emission pattern is expected if 
massive stars are the dominant source of galactic \al\ (e.g.~Diehl et 
al.~1995).
In order to understand the 1.8 \MeV\ sky globally, considerable 
effort has been made in modelling the COMPTEL \al\ data
(Chen et al.~1995; Diehl et al.~1995, 1996, 1997; 
Kn\"odlseder et al.~1996a).
Since imaging results suggest that massive stars are among the most 
plausible sources of \al, the modelling focused primarily on finding 
optimum tracers of massive stars in the Galaxy.
As such, molecular gas models, based on the CO survey of Dame et 
al.~(1987), and far-
\newpage \noindent
infrared maps, obtained by the DIRBE telescope aboard 
the {\em Cosmic Background Explorer} ({\em COBE}) provide reasonable first 
order descriptions of the data, but significant residual emission remains 
towards Cygnus, Vela, and Carina when the maps are fitted to the data.
Additionally, the intensity contrasts between positive and negative 
longitudes or the inner and outer Galaxy for these maps differ from 
those of the 1.8 \MeV\ data.

Advancing the search for a satisfactory tracer of 1.809 \MeV\ emission, we 
extend in this paper our set of possible tracers to an extensive 
database of all-sky maps, covering the entire explored wavelength range from 
the $\lambda \sim 10$ m radio band up to the energetic $E > 100$ \MeV\ 
\gray\ photons.
Since the Galaxy is transparent to \gray s, each all-sky map of our 
database represents a specific hypothesis about the galactic 
distribution of \al.
Hence, the results of the comparison of these maps to COMPTEL 1.8 \MeV\ 
data may be interpreted in terms of these hypotheses, providing physical 
insight into the origin of galactic \al.
The database and preparation of the all-sky maps is discussed in detail 
in a separate paper by Kn\"odlseder et al.~(1998a) which gives also more 
details on the analysis methods used for the study.
Here we update sky maps for the FIRAS far-infrared line 
emission at 158 $\um$ (C$^+$) and 205 $\um$ (N$^+$) which have 
become available recently (Pass 4 data products).

\bsk
\ni 2. TRACER MAP COMPARISON
\ssk
\ni
\label{sec:comparison}

For the comparison, the 31 all-sky maps of the database have been convolved 
with the 3-dimensional COMPTEL point spread function to provide model 
distributions of 1.8 \MeV\ source counts in the COMPTEL imaging data space.
These source models are then fitted along with a model for the 
instrumental background contribution to COMPTEL data in the energy 
interval $1.7-1.9$ \MeV.
The background model is based on the event distribution in adjacent 
energy intervals and has been proven to suppress most of the continuum 
emission in the data analysis (Kn\"odlseder et al.~1996b).
In order to reduce systematic uncertainties in the analysis the $\phibar$ 
distribution of the background model is adjusted by the fit
(for an analysis using an alternative background model see Bloemen et 
al., these proceedings).
After fitting, a residual analysis is performed for each map using the 
`software collimation' technique (Diehl et al.~1993) and the maximum 
likelihood ratio test (de Boer et al.~1992).
The first method provides longitude profiles of the residual emission 
in the raw data space while the second method results in significance 
maps of residual point source emission.
It turned out that this residual analysis is less sensitive to 
systematic uncertainties in our background modelling procedure than
the likelihood analysis used previously (Kn\"odlseder et al.~1998a).

\begin{figure}
\centerline{\psfig{file=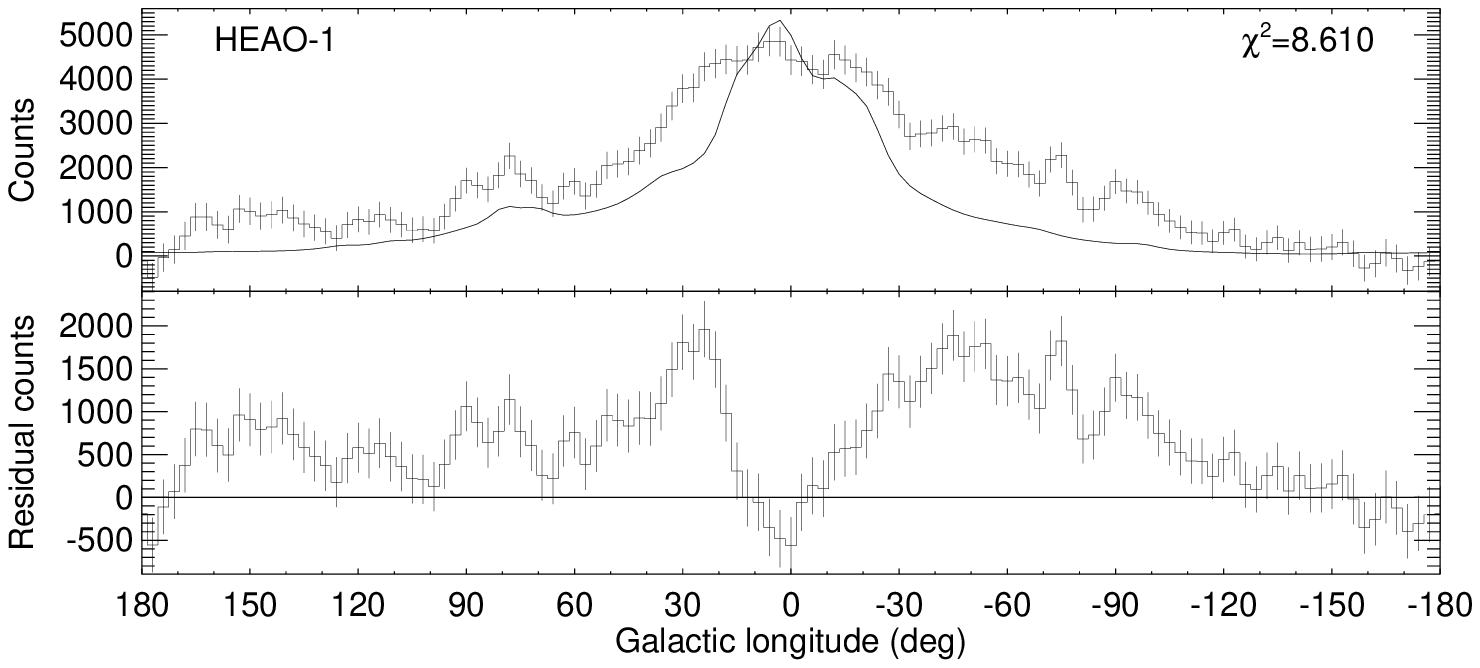, width=6.3cm}
            \hfill
            \psfig{file=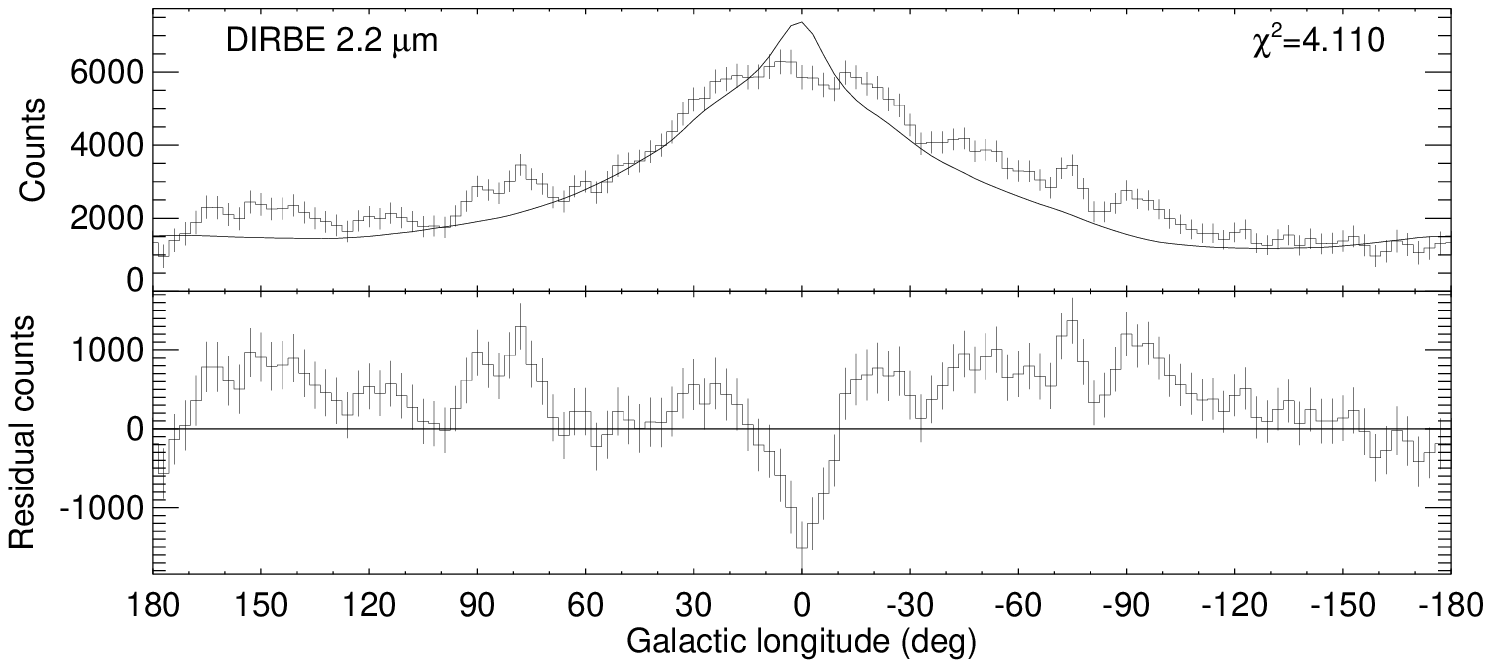, width=6.3cm}}
\centerline{\psfig{file=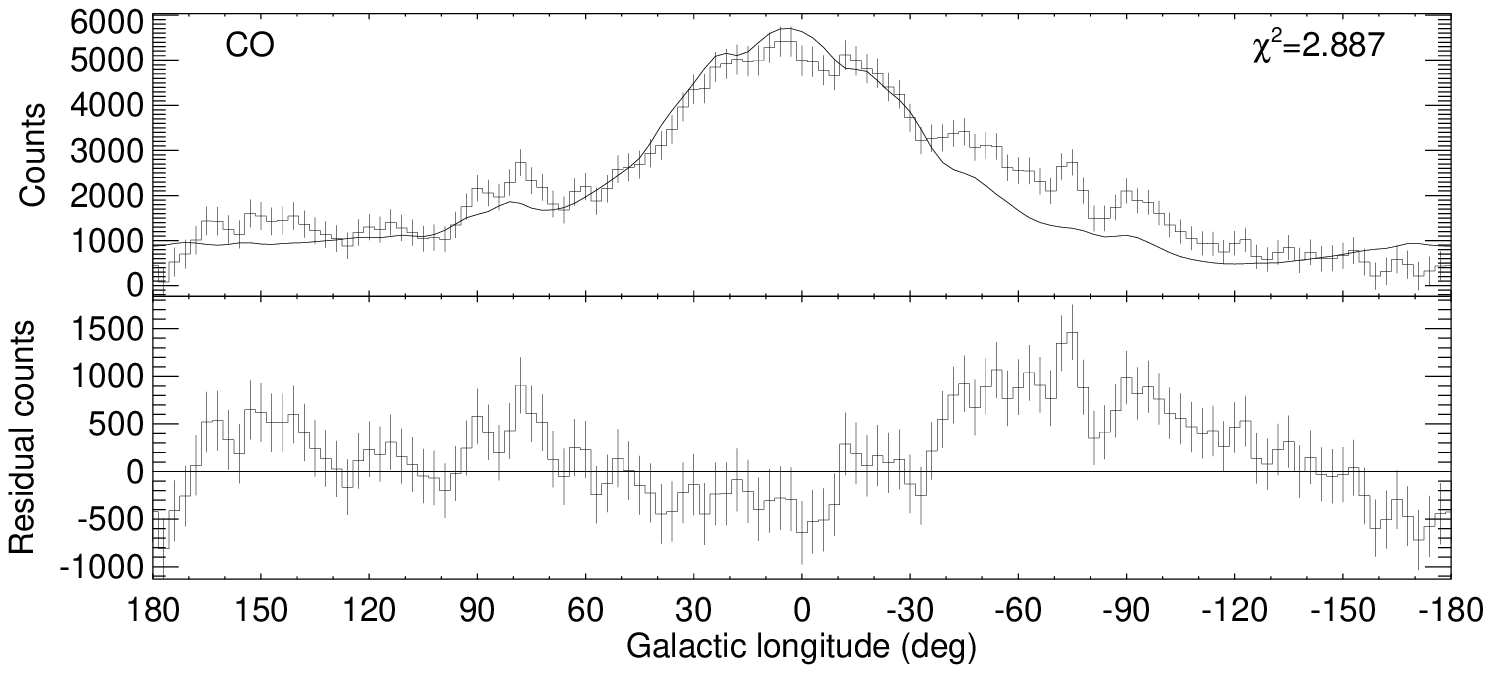, width=6.3cm}
            \hfill
            \psfig{file=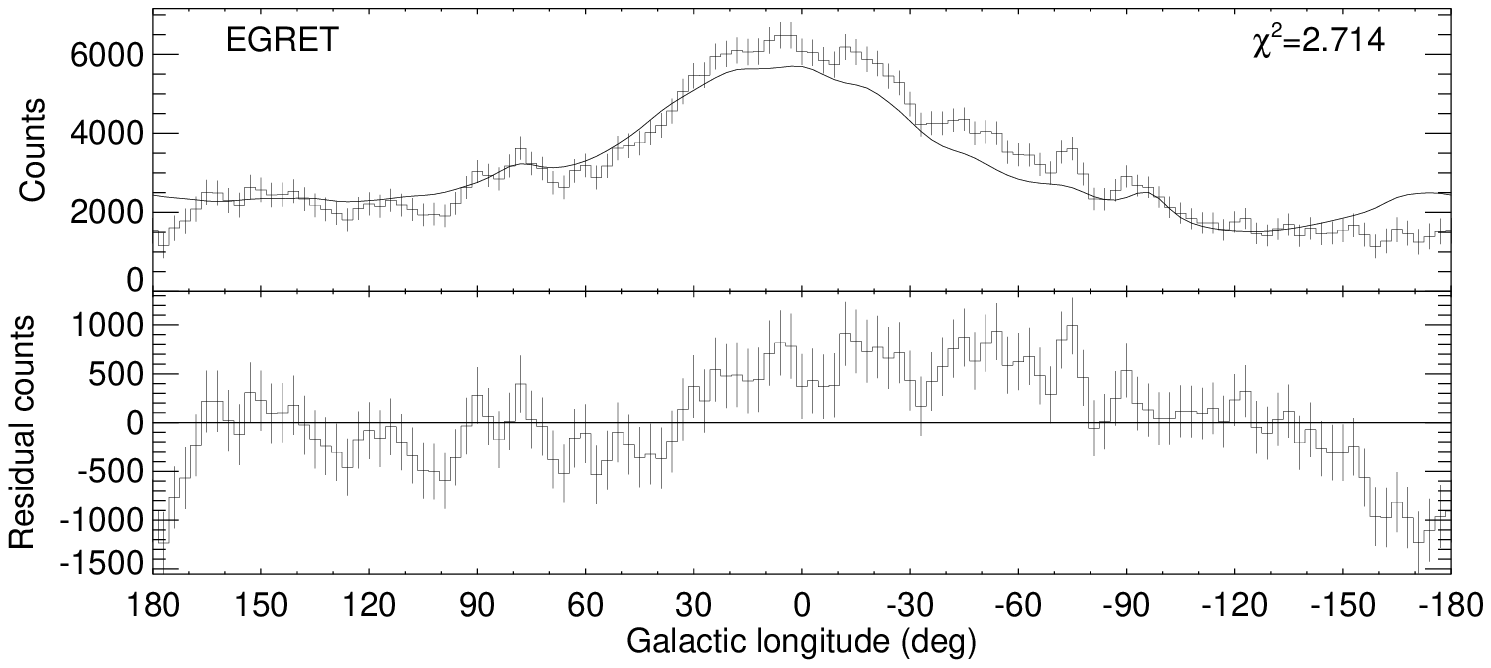, width=6.3cm}}
\centerline{\psfig{file=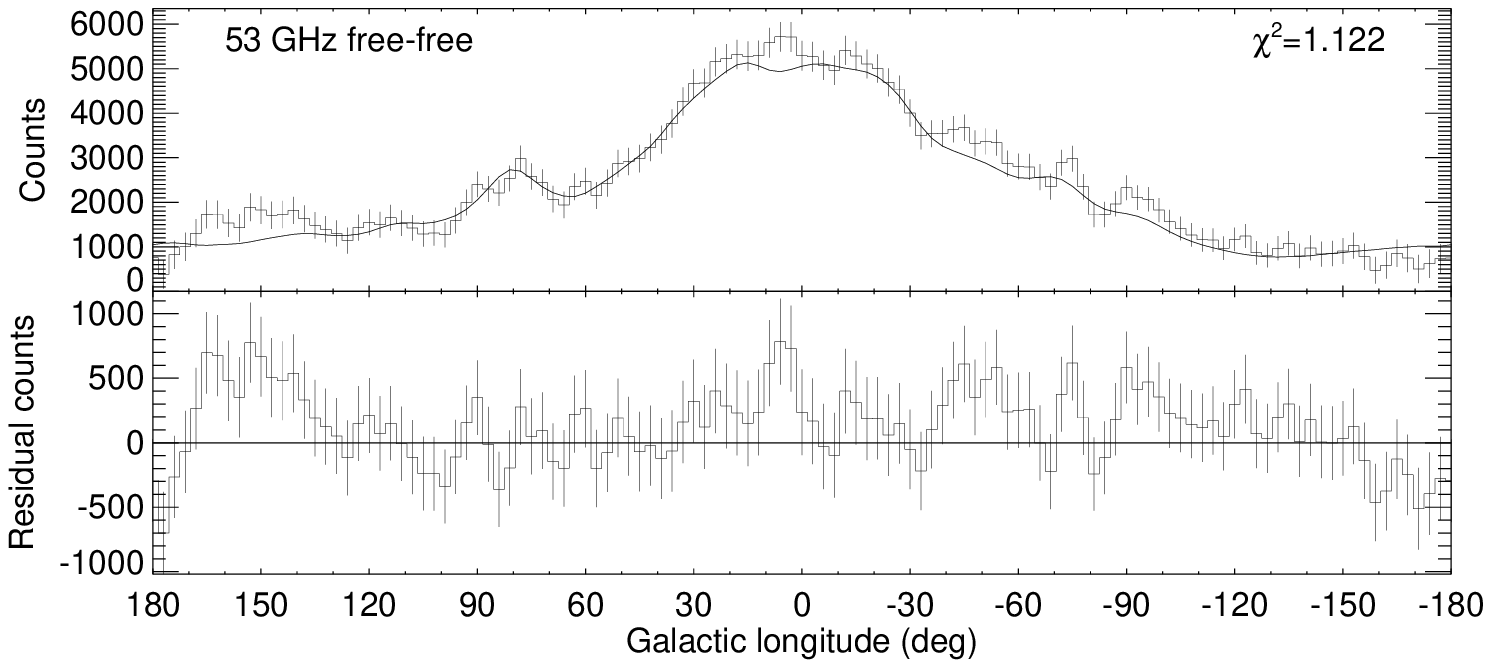, width=6.3cm}
            \hfill
            \psfig{file=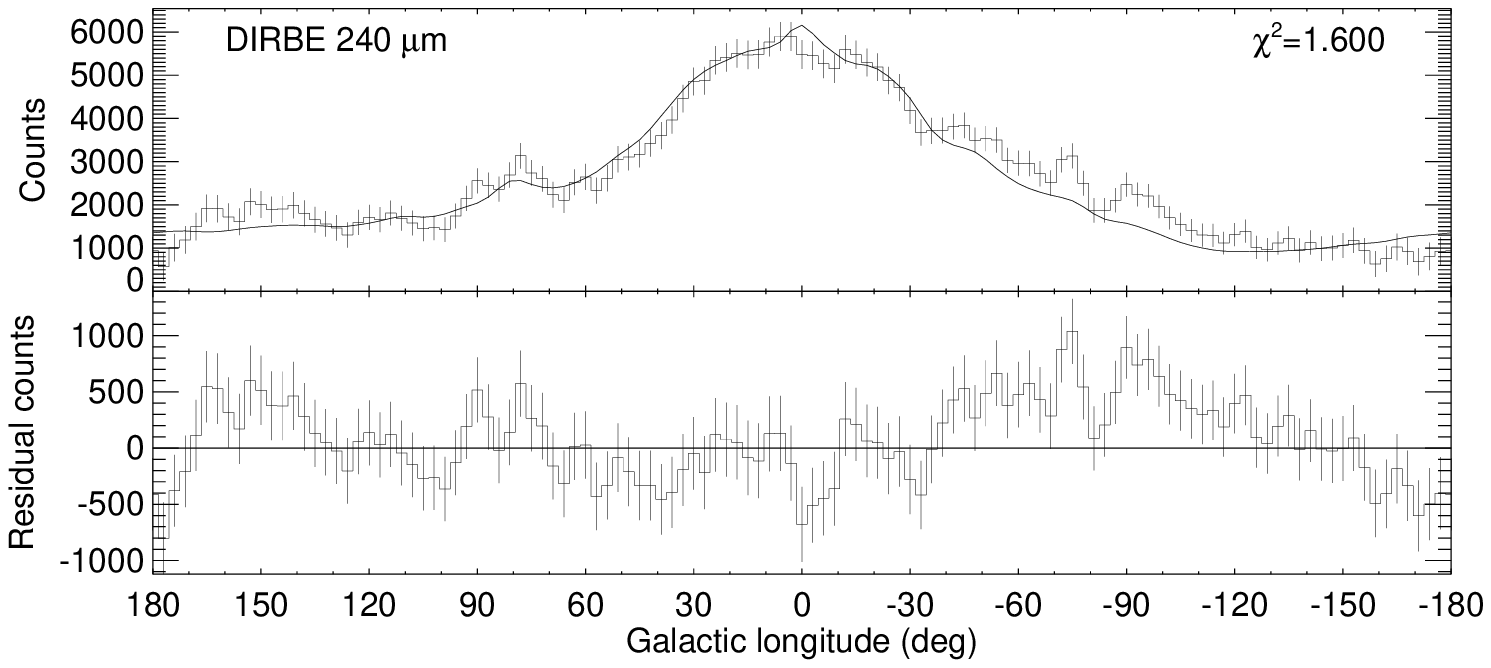, width=6.3cm}}
\centerline{\psfig{file=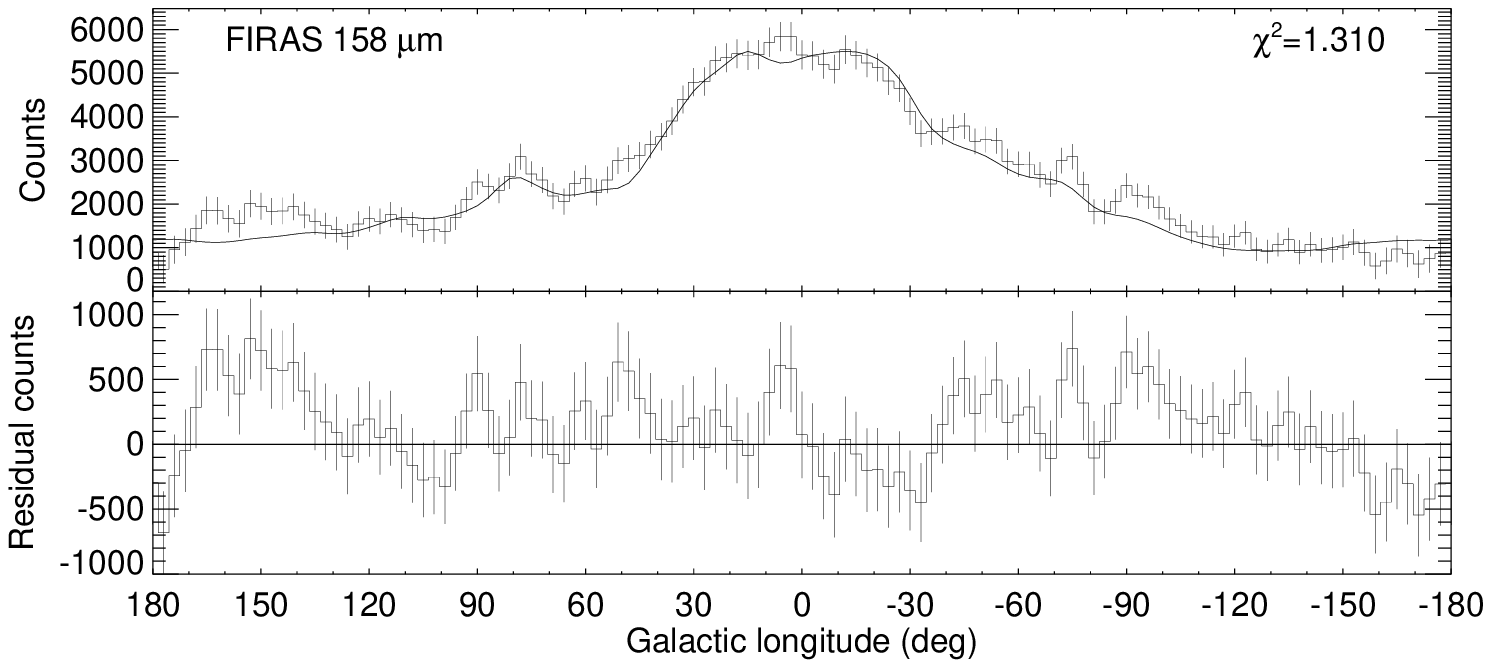, width=6.3cm}
            \hfill
            \psfig{file=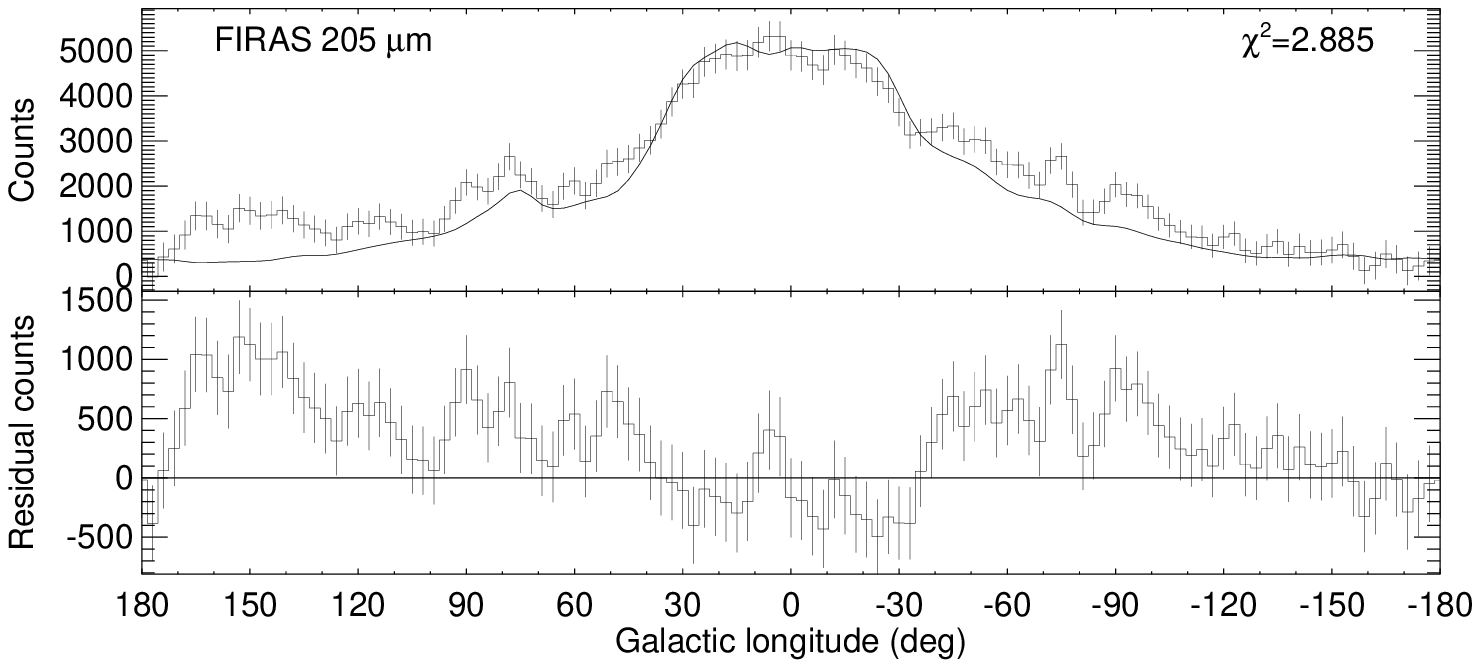, width=6.3cm}}
\centerline{\psfig{file=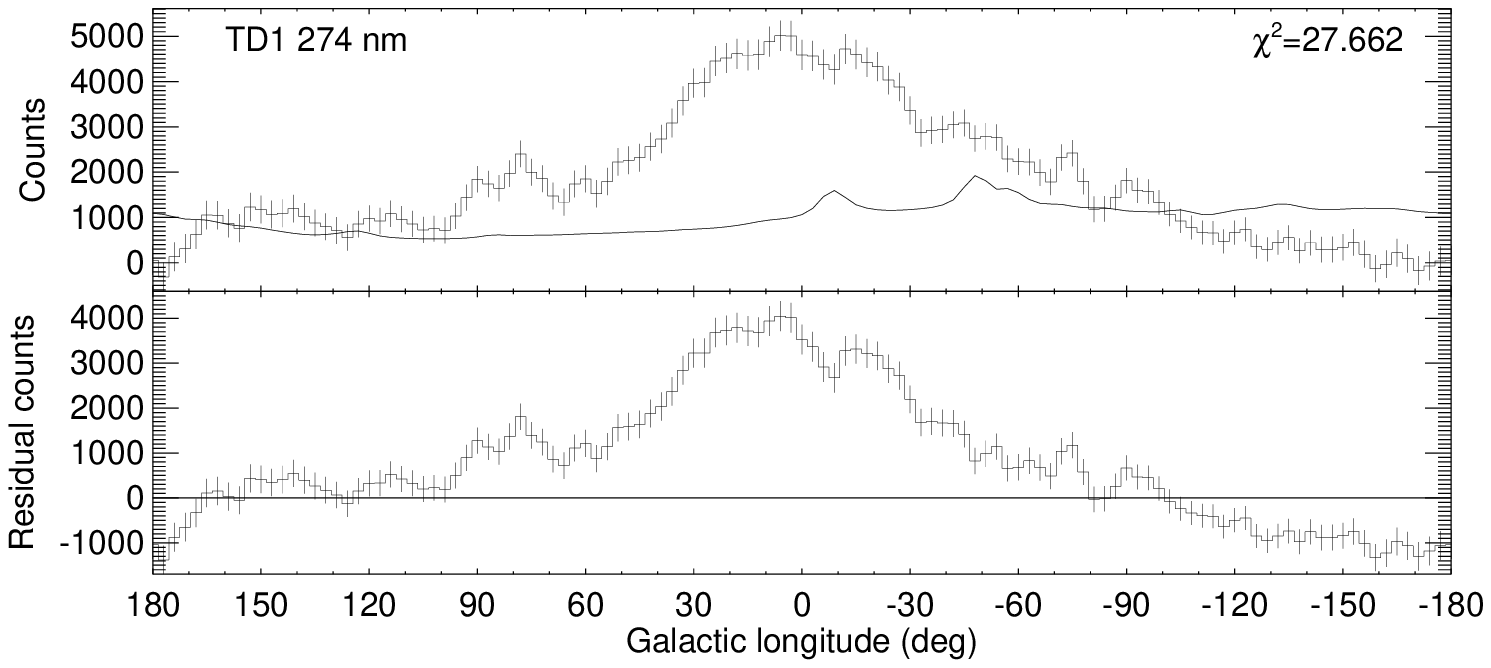, width=6.3cm}
            \hfill
            \psfig{file=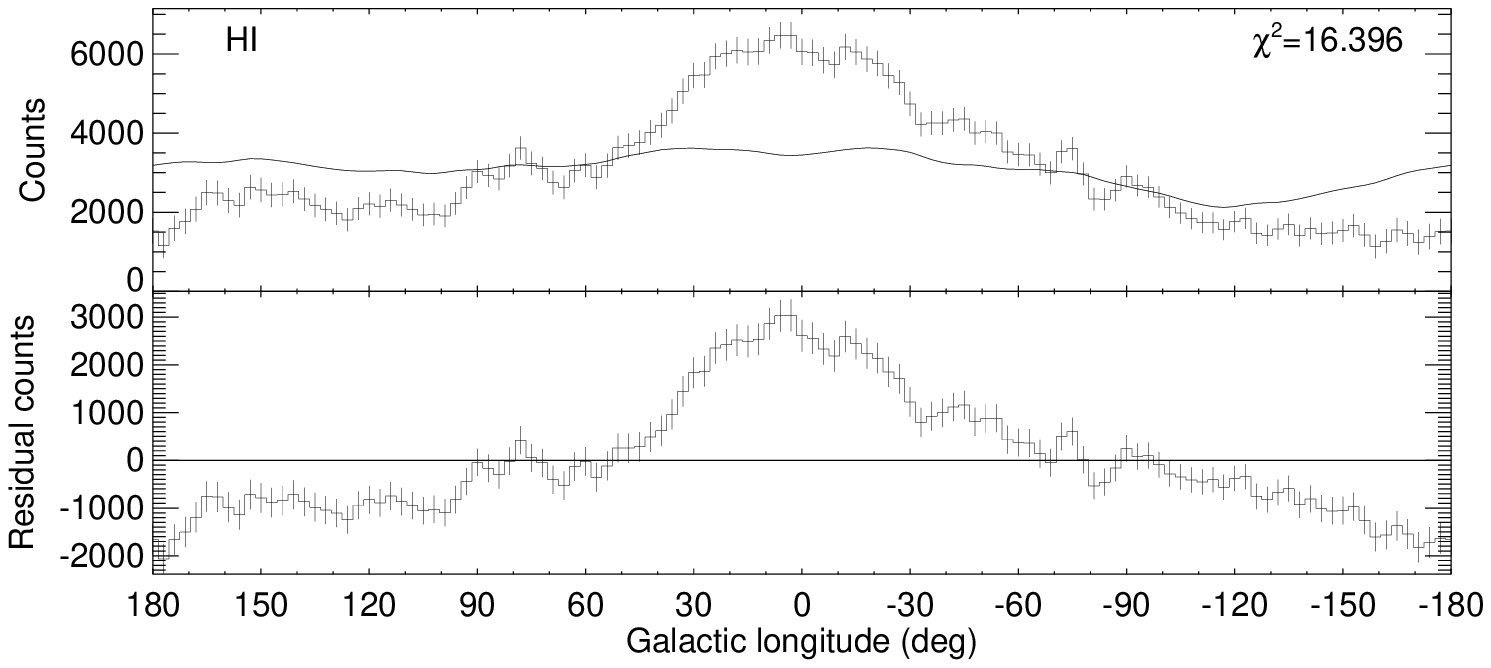, width=6.3cm}}
    \caption{\label{fig:residuals}FIGURE 1.
    Software collimator longitude scans for some of the all-sky maps.
    The histograms show the background subtracted $1.8$ MeV counts, 
    the solid lines show the corresponding all-sky map profiles 
    convolved into the COMPTEL imaging data space.
    The reduced $\chi^2$ values for the residual longitude profiles 
    are quoted as measures of the correlation between the $1.8$ MeV 
    and the all-sky map longitude profiles.
    } 
\end{figure}

\bsk
\ni
2.1 Old Stellar Population
\ssk
\ni

Maps in our database which are related to the old stellar population 
are the HEAO-1 hard X-ray map and the DIRBE near infrared maps at 
$\lambda \le 4.9$ $\um$.
The X-ray map traces the population of X-ray binaries which shows a 
strong concentration towards the inner Galaxy.
The longitude profile for the HEAO-1 map (cf.~Fig.~1) 
clearly demonstrates that this concentration is much stronger than that 
of 1.8 \MeV\ emission.
When fitted to the data, the hard X-ray maps leave significant 1.8 
\MeV\ residual emission in the outer disk regime.
The near infrared maps, tracing the population of K and M giants, can not 
explain the 1.8 \MeV\ data either (as an example the longitude profile 
for the DIRBE 2.2 $\um$ map is shown in Fig.~1).
Obviously, this population obeys a rather smooth and symmetric profile, in 
contrast to the structured and asymmetric distribution of the 1.8 
\MeV\ emission.
Indeed, the population of K and M giants is usually modelled by a 
superposition of a probably bar-shaped bulge and an exponential disk 
with radial scale length between $1-3$ kpc (Wainscoat et al.~1992), 
while fitting geometrical models to COMPTEL $1.8$ MeV data suggests a 
larger radial scale length of $4.5\pm0.4$ kpc for the galactic 
distribution of \al\ and no evidence for a galactic bulge component
(Diehl et al.~1998).

\bsk
\ni
2.2 Young Stellar Population
\ssk
\ni

There are a number of maps in our database which are related to the 
massive, hence young, stellar population, although the relation may 
be quite convoluted.
Molecular gas as measured in CO traces the raw material of star 
formation, hence it maps the potential for star formation throughout 
the Galaxy.
Microwave data can be used to extract the distribution of galactic 
free-free emission arising from photoionization of the interstellar 
medium (ISM) by massive stars.
Far-infrared continuum emission reveals the distribution of interstellar 
dust which has been shown to correlate with galactic \HII\ regions in the 
galactic plane (Broadbent et al.~1989).
Far-infrared line emission traces neutral and ionized gas cooling, 
related to massive star heating of the ISM by FUV photons.
High-energy \gray s arise from the interaction of cosmic-ray nuclei 
with the interstellar gas and show a distribution similar to that of 
CO (the high-energy \gray\ longitude profile as observed by COS-B has been 
often used as a template to determine the galactic 1.809 \MeV\ flux).

Figure 1 illustrates that all these maps provide a 
reasonable first order description of the COMPTEL 1.8 \MeV\ data.
But most of them can not explain some peculiar features of the data, 
like the relative large 1.8 \MeV\ emission in the southern 
hemisphere ($l<0\deg$) or the emission peaks towards Cygnus 
($l\approx80\deg$), Carina ($l\approx-75\deg$), and Vela
($l\approx-90\deg$).
The only maps in the database which apparently explain all of the 
features of the 1.8 \MeV\ data are the 53 GHz free-free emission map 
and the FIRAS 158 \um\ line-emission map.
The first map was derived by Bennett et al.~(1992) from microwave data 
taken by DMR aboard {\em COBE} using the different spatial and spectral 
morphology of the dominant emission processes at this frequency.
Since the Galaxy is transparent at microwave frequencies, the map traces 
the distribution of ionized matter throughout the entire Galaxy.
The second map was obtained using the FIRAS telescope aboard the {\em 
COBE} satellite.
The 158 \um\ line arises from a fine-structure ground state transition 
of C$^+$, and is considered as the dominant coolant of neutral 
interstellar gas (Tielens \& Hollenbach 1985).
Galactic absorption is also negligible at this wavelength.
Remaining weak residuals have to be studied carefully, as they may be 
related to systematic uncertainties in our instrumental background 
model.
For example, significant residuals near the anticenter and towards 
the southern galactic pole area have been found to show no 1.8 \MeV\ 
line feature in the spectral analysis (Kn\"odlseder et al.~1998a).

The fitting procedure provides scaling factors for the tracer maps 
which allow to determine the ratio between 1.8 \MeV\ intensities and 
free-free and 158 \um\ intensities, respectively.
For the intensity ratio between 1.8 \MeV\ and 53 GHz free-free 
emission we obtain 
$0.86\pm0.03$ ph cm$^{-2}$ s$^{-1}$ sr$^{-1}$ K$^{-1}$
(free-free intensities are usually quoted in units of antenna 
temperature).
The intensity ratio between 1.8 \MeV\ and 158 \um\ line emission was 
determined to $12.0\pm0.4$ ph erg$^{-1}$
(158 \um\ line intensities are usually given in units of
erg cm$^{-2}$ s$^{-1}$ sr$^{-1}$).
From these ratios we determine the inner radian 1.8 \MeV\ flux by 
integration over $-30\deg < l < 30\deg$ and $|b|<10\deg$, resulting 
in $(2.6\pm0.1)\,10^{-4}$ ph cm$^{-2}$ s$^{-1}$ for both maps.
This is consistent with flux determinations from image reconstruction 
using the same instrumental background model (Oberlack et al.~1998), but 
significantly smaller than the value found by Bloemen et al.~(these 
proceedings) using a different background model.
The discrepancy reflects our systematic uncertainties due to the 
instrumental background modelling procedure, which is still 
limiting our analysis accuracy.

\bsk
\ni
2.3 Gould's Belt
\ssk
\ni

The local population of massive stars forms an inclined system with 
respect to the galactic plane, referred to as Gould's Belt.
It is dominated by a few associations, predominantly in the third 
and fourth quadrants of galactic longitude.
Recent analysis based on Hipparcos data has revealed an age of 
$30-40$ Myr for the system and distances between 400 and 600 pc to the 
individual members (Lindblad et al.~1997; Torra et al.~1997).
It has been suggested earlier that nucleosynthesis activity of the 
massive stars in this system could lead to a distinct 1.8 \MeV\ emission 
feature on the sky (Diehl et al.~1996, see also Diehl et al., these 
proceedings).
Fitting of COMPTEL 1.8 \MeV\ data using geometrical models for the 
Gould's Belt population, however, showed no significant correlation
(Diehl et al.~1997).

Our database contains 4 all-sky maps obtained by the TD1 satellite in 
the UV waveband which are dominated by local O and B stars, and which 
nicely delineate Gould's Belt.
Comparison of these maps to our data reveals that the bulk of 1.809 \MeV\ 
emission is certainly not correlated to this structure 
(cf.~Fig.~1).

\bsk
\ni 3. DISCUSSION
\ssk
\ni
\label{sec:discussion}

Our multi-wavelength tracer map comparison clearly demonstrates that
the 1.809 \MeV\ intensity distribution does not correlate with the 
old stellar population.
Proposed \al\ candidate sources among these objects are novae and 
low-mass AGB stars, and we conclude that those can be excluded as prolific 
sources of \al.
Apparently, the 1.809 \MeV\ emission shows clearly the characteristics 
of a young stellar population which is an asymmetric emission profile 
along the galactic plane with regions of enhanced intensity due to 
galactic spiral structure and clustered massive star formation.
It is remarkably, however, that COMPTEL 1.8 \MeV\ data are best 
explained by tracers of the extreme Population I, i.e.~tracers of 
very massive stars.
The 53 GHz free-free emission map traces the distribution of the 
ionized interstellar medium which is dominantly ionized by UV photons 
of stars with initial masses above $\sim20\Msol$ (Kn\"odlseder 1998b; 
see also Kn\"odlseder, these proceedings).
The 158 \um\ \CII\ line has been shown to be an excellent tracer of OB 
star formation activity arising from the interface of \HII\ regions 
and molecular clouds (Stacey et al.~1991).
The excellent correlation of COMPTEL 1.8 \MeV\ data with these two 
tracer maps strongly suggests that \al\ is also produced by very 
massive stars, making core-collapse supernovae and/or Wolf-Rayet stars 
the primary candidate sources.
Small \al\ contributions from AGB stars or novae may well exist, but 
they are not required by the present data.

\bsk
\baselineskip = 12pt
{\abstract \ni ACKNOWLEDGMENTS
It is a pleasure to thank M.~Leising for helpful discussions and comments.
The COMPTEL project is supported by the German government through
DARA grant 50 QV 90968, by NASA under contract NAS5-26645, and by
the Netherlands Organisation for Scientific Research NWO.
}
\bsk
\baselineskip = 12pt

{\references \ni REFERENCES
\ssk

\ref Bennett, C.~L., Smoot, G.~F., Hinshaw, G., et al.~1992, 
     \apj, 396, L7
\ref  Broadbent, A., Haslam, C.~G.~T., \& Osborne, J.~L.~1989, 
      \mnras, 237, 381
\ref  Chen, W., Gehrels, N., \& Diehl, R.~1995, \apj, 440, L57
\ref  Dame, T.~M., et al.~1987, \apj, 322, 706
\ref  de Boer, H., et al.~1992, in: Data Analysis in Astronomy - 
      IV, ed.~D.~Ges\`u, 241 (Plenum Press)
\ref  Diehl, R., et al.~1993, \aaps, 97, 181
\ref  Diehl, R., et al.~1995a, \aap, 298, 445
\ref  Diehl, R., et al.~1996, \aaps, 120C, 321
\ref  Diehl, R., et al.~1997, in Proc.~of the 4th Compton 
      Symposium, AIP 410, 1114
\ref  Diehl, R., et al.~1998, in preparation for \aap
\ref  Kn\"odlseder, J., et al.~1996a, \aaps, 120C, 335
\ref  Kn\"odlseder, J., et al.~1996b, Proc.~SPIE 2806, 386
\ref  Kn\"odlseder, J., et al.~1998a, \aap, submitted
\ref  Kn\"odlseder, J.~1998b, \apj, in press
\ref  Lindblad, P.~O., et al.~1997, ESA SP-402, 507
\ref  Oberlack, U., et al.~1996, \aaps, 120C, 311
\ref  Oberlack, U., et al.~1998, in preparation for \aap
\ref  Wainscoat, R.~J., et al.~1992, \apjs, 83, 111
\ref  Stacey, G.~J., et al.~1991, \apj, 373, 423
\ref  Tielens, A.~G.~G.~M., \& Hollenbach, D.~1985, \apj, 291, 722
\ref  Torra, J., et al.~1997, ESA SP-402, 513

}                      

\end{document}